\documentclass[12pt]{iopart}
\usepackage{cite}
\usepackage{epsf}
\usepackage{epsfig}
\usepackage{float}
\usepackage{amssymb}
\usepackage{amsfonts}
\usepackage{color}

\def\ep{{\epsilon}}

\def\k{{{\bf k}}}
\def\om{{\omega}}
\def\omp{{\omega^\prime}}

\def\nnu{{\nonumber}}

\def\g{{\bf{g}}}

\def\P{{\bf{P}}}

\def\beq{\begin{equation}}
\def\eeq{\end{equation}}
\def\beqa{\begin{eqnarray}}
\def\eeqa{\end{eqnarray}}

\def\g0{{\gamma_0}}

\def\Re{{\mbox{Re}}}
\def\Im{{\mbox{Im}}}

\def\prb{Phys.\ Rev.\ B }
\def\prl{Phys.\ Rev.\ Lett.\ }

\def\JPCM{J.\ Phys.\ Condens.\ Matter }
\def\JMMM{J.\ Mag.\ Mag.\ Materials.\ }
\def\etal{{\it et.\ al} }

\def\RMP{Rev.\ Mod.\ Phys.\ }
\def\JPSJ{J.\ Phys.\ Soc.\ Japan}
\def\EPL{ Europhys.\  Lett.\ }

\newcommand\bear{\begin{eqnarray}}
\newcommand\eear{\end{eqnarray}}
\newcommand\bea{\begin{align}}
\newcommand\ena{\end{align}}
\newcommand{\field}[1]{\mathbb{#1}} 

\begin{document}

\title{From mixed valence to the Kondo lattice regime}

\author{Pramod Kumar and N.\ S.\ Vidhyadhiraja}
\address{Theoretical Sciences Unit, \\ Jawaharlal Nehru Centre for Advanced
Scientific Research,\\ Bengaluru 560064, India.}

\begin{abstract}
Many heavy fermion materials are known to crossover from the Kondo lattice
regime to the mixed-valent regime or vice-versa as a function of pressure or 
doping. 
We study this
crossover theoretically by employing the periodic Anderson model within the
framework of the dynamical mean field theory. Changes occurring
in the dynamics and transport across this crossover are highlighted. 
 As the valence is decreased (increased) relative to the Kondo lattice regime, 
the Kondo resonance broadens significantly, 
while the lower (upper) Hubbard band moves closer to the Fermi level. 
The resistivity develops a two peak structure in the mixed valent regime: a 
low temperature
coherence peak and a high temperature `Hubbard band' peak.
These two peaks merge yielding a broad shallow maximum upon decreasing the
valence further.
The optical conductivity, likewise exhibits an unusual absorption
feature (shoulder) in the deep mid-infrared region, which grows in intensity 
with decreasing valence. The involvement of the Hubbard bands in dc transport,
and of the effective $f$-level in the optical conductivity are shown to be
responsible for the anomalous transport properties.  A two-band 
hybridization-gap model, which neglects incoherent effects due to many-body
scattering, commonly employed to understand the optical response
in these materials is shown to be inadequate, especially in the mixed-valent
regime. Comparison of theory with experiment carried out for (a) dc resistivities of CeRhIn$_5$,
Ce$_2$Ni$_3$Si$_5$, CeFeGe$_3$ and YbIr$_2$Si$_2$; (b) pressure dependent resistivity
of YbInAu$_2$ and CeCu$_6$; and (c) optical conductivity measurements in YbIr$_2$Si$_2$ yields 
excellent agreement.
\end{abstract}

\pacs{71.27.+a Strongly correlated electron systems; heavy fermions - 
75.20.Hr Local moment in compounds and alloys; Kondo effect, valence
fluctuations, heavy fermions}

\submitto{\JPCM}

\maketitle

\section{Introduction}
\label{sec:intro}

Rare earth lanthanides and actinides~\cite{grew91,wach} exhibit a wide range of behaviour
such as heavy fermions (HFs), mixed valence (MV), proximity of 
superconductivity and 
magnetism, quantum critical points etc. Such behaviour arises through an 
interplay of a variety of factors such as hybridization between conduction
bands and deep $f$-levels, orbital degeneracy, crystal field 
effects, long range spin interactions and most importantly local Coulomb 
repulsion~\cite{varm,aepp,hews,fisk,degi}. 
   
   In this work, we present a detailed and systematic theoretical
investigation of a regime that borders on heavy fermions at one end 
and on the mixed valence regime at the other.
Experimentally, such a crossover from HFs to MV or vice-versa has been
 observed to happen through 
pressure or doping~\cite{jacc,ocko,baue,kaga}.  The effects of such
 a crossover have 
been investigated for several materials~\cite{taku,yuan,kneb,mats,
kacz01,adro,kacz02,sere,fran01,fran02}. For example, the first known
heavy fermion superconductor
CeCu$_2$Si$_2$~\cite{ocko}, when doped with Yttrium (a non-magnetic homologue
 of Ce) shows increasing mixed valent character.
The resistivity exhibits a two peak structure in the temperature range 2K to 300K.
Varying doping concentration results in gradual coalescing of the two peak
structure into a single broad peak.
For the heavy fermion compound YbCu$_4$Ag~\cite{baue},
the ambient pressure resistivity as a function of temperature shows a broad peak 
which is characteristic of mixed valent compounds. By applying pressure, the broad peak
sharpens and the peak position also shifts to lower values of $T$, thus indicating
a crossover to the Kondo lattice regime. The $T^2$ coefficient of the low temperature
Fermi-liquid resistivity also increases sharply with pressure, implying a decrease
in the coherence scale.
CeBe$_{13}$~\cite{kaga} also shows a pressure-induced crossover from Kondo lattice regime to 
mixed-valent regime, as seen in the changes in the coherence peak in the resistivity. The trend
in this material is opposite to that seen in the previous example, YbCu$_4$Ag.~\cite{baue}

Most previous theoretical attempts
to describe the effects of pressure on heavy fermion materials 
have employed the single-impurity Anderson model~\cite{leen,grew88,cox}. 
An illustrative and important work in 
this context is that of Chandran et al~\cite{leen}.
They use a phenomenological model comprising
a competition between elastic energy cost and valence fluctuations
induced magnetic energy gain. The magnetic energy for an Anderson lattice
model was computed by using the free energy of an impurity Anderson model
and ignoring lattice coherence effects. The free energy itself was arrived
at through a slave-boson mean field approximation~\cite{pier01} which is a 
static approximation and is thus unable to treat dynamical effects of valence 
fluctuations. Phenomenological expressions were used to model volume dependence
of the parameters, and the pressure was obtained by using $P(T,V)=-\partial F/\partial V$.
The authors were able to describe continuous and discontinuous valence transitions
in a single framework.
However, since the impurity Anderson model was used, lattice coherence
effects were ignored. Transport quantities were
not calculated. Recent years have seen the use of lattice models to describe
the concentrated Kondo systems.
 The minimalist model that accounts for a large
part of heavy fermion and mixed-valent behaviour is the periodic 
Anderson model (PAM) which
represents a lattice of localized $f$-orbitals with a Hubbard repulsion $U$
hybridizing locally with a wide non-interacting conduction band.
The PAM has of course been explored extensively using a wide 
range of methods and techniques such as dynamical mean field theory (DMFT)~\cite{geor,prus95},
extended DMFT~\cite{qimi}, cluster expansions~\cite{maie}, finite size simulations~\cite{scal} etc.
Within the framework of dynamical mean field theory (DMFT)~\cite{geor,prus95}, the self-energy
 becomes
purely local or momentum independent, which simplifies the problem 
significantly, while retaining the competition between intinerancy and 
localization in a non-trivial way. The momentum independence of the self--energy
implies that we can view lattice problems as locally self-consistent 
impurity problems, hence the solution of PAM, for instance reduces to
that of a single-impurity Anderson model with a self-consistent 
hybridization. The various impurity solvers that have been adapted
to solve the effective impurity problem are numerical renormalization
group~\cite{prus00}, quantum Monte Carlo~\cite{jarr95}, 
exact diagonalization~\cite{roze95}, perturbation theory
methods such as iterated perturbation theory~\cite{vidh00}, 
local moment approaches~\cite{vidh04},
and slave-particle approaches~\cite{sun} etc. 

The Kondo lattice regime within the PAM has been investigated heavily by us~\cite{vidh04,vidh05,vidh06} and the other groups~\cite{prus00,glos,psun}
within dynamical  mean field theory and cluster extensions ignoring the d$-$f repulsion effects. 
We have used local moment approach(LMA) within the DMFT framework previously to understand the PAM in 
the Kondo lattice (KL)limit~\cite{vidh05}. The LMA is a non-perturbative diagrammatic theory based approach 
developed by Logan and co-workers~\cite{loga98}. This approach has been benchmarked extensively against methods such as
Bethe Ansatz~\cite{loga98} and numerical renormalization group~\cite{bull}. 
Excellent quantitative agreement has been 
found, thus providing justification for its use as an impurity solver within DMFT. 
Further advantages of using LMA are, that real frequency 
quantities are obtained directly with reasonable computation expense at all temperature and interaction strengths. 
Our focus in the LMA$+$DMFT approach to PAM was on universality and scaling in dynamics and transport.
 A single low energy scale was found to characterize the spectra and
transport in the Kondo lattice regime~\cite{vidh04}. The mixed-valent regime
was not our focus, nevertheless, our studies indicated the absence of universality and scaling,
although adiabatic continuity to the non-interacting limit was seen. 
The dc and optical transport properties were compared to several
Kondo insulators and heavy fermion metals and excellent quantitative 
agreement was found~\cite{vidh03,vidh06}. A few mixed-valent
materials such as YbAl$_3$ and the skutterudite compound CeOs$_4$Sb$_{12}$ 
were also considered ~\cite{vidh07} and again
good agreement between theory and experiment was found. In a recent work, valence transition
in Ytterbium and Europium intermetallics was studied by Zlatic and Freerics ~\cite{zlat} employing a multi-component
Falicov-Kimball (FK) model within dynamical mean field theory. 
The authors argue that a complete description would entail a solution of the periodic Anderson model
combined with the FK model. However, since this is challenging, they choose 
to solve just the FK model, albeit a multi-component one. Transport quantities like
dc and optical conductivity, thermopower and magnetoresistance etc
were calculated and qualitatively compared to the experiment. Using an equation
of motion decoupling approximation, Bennard and Coqblin~\cite{coqb} have investigated the PAM
and explored the variation of the $f$-valence with various parameters of the model. 
They use the results of this study to understand pressure dependent valence changes. 
Miyake and co-workers~\cite{wata} have used a variety of methods including the slave-boson approximation, 
the density matrix renormalization group etc to investigate the one dimensional extended PAM (EPAM),
 which includes the $d-f$ repulsion effects represented by $U_{cf}$. The EPAM has been investigated
within dynamical mean field theory by Sugibayashi and Hirashima ~\cite{sugi}using quantum Monte-Carlo methods.
Their focus has been to understand the interplay between U$_{cf}$, valence fluctuations and superconductivity.
  In a recent work Ylvisaker~\cite{ylvi} \etal 
combined local density approximation with dynamical mean field theory to understand the valence
fluctuation and the valence transition in the Yb metal. For the impurity solver, they have used 
Hirsh-Fye quantum Monte-Carlo and continuous time   quantum Monte-Carlo.
 They reproduce the experimentally observed valence transition, and conclude that
the Yb metal is a fluctuating valence material rather than an intermediate-valent one. 
As mentioned before and as illustrated through the above
mentioned studies, there has been substantial work on the valence transition and its effect
on the spectral quantities within the PAM, however the effects on transport quantities
due to valence fluctuations and the crossover regime between the Kondo lattice and the mixed-valent regime
have received scant attention.

In this work, 
we focus on such a KL-MV crossover using the LMA+DMFT approach to the PAM.
We assume that effects of pressure/doping would be to change the model parameters,
and hence a scan of the parameter space within the PAM framework should be able to
provide insight into the crossover regime.
We highlight the changes occuring in the dynamics and transport properties
as a result of this crossover. We find several new results such as a two peak
resistivity and anomalous absorption features in the optical conductivity in certain parameter regimes.
We provide theoretical explanations for these anomalies, and show that
such behaviour does indeed exist in real materials and may be explained 
quantitatively using the present approach. The paper is structured as follows:
We present the model and formalism in the next section. The results
and discussions are in sections~\ref{sec:res} and ~\ref{sec:disc} respectively. A comprehensive
range of experimental measurements is shown to be described by our theoetical 
results in section ~\ref{sec:comp}. We conclude in 
section~\ref{sec:conc}.

\section{Model and Formalism}

The periodic Anderson model is one of the simplest models representing
a paradigm for understanding the physics of heavy-fermion compounds. 
In standard notation, the Hamiltonian for PAM is given by

\bear
\fl \hat{H} =  \epsilon_c \sum_{i\sigma} c_{i\sigma}^\dag 
c^{\phantom{\dag}}_{i\sigma}\, -t\!\!\sum_{(i,j),\sigma} ( c_{i\sigma}^\dag 
c^{\phantom{\dag}}_{j\sigma} + {\rm h.c.})
 +V \sum_{i\sigma}(f_{i\sigma}^\dag c^{\phantom{\dag}}_{i\sigma}+{\rm h.c.})
\nnu \\
+ \sum_{i\sigma}\left(\epsilon_f +\frac{U}{2} f_{i,-\sigma}^\dag 
f^{\phantom{\dag}}_{i,-\sigma}\right)f_{i,\sigma}^\dag f^{\phantom{\dag}}_{i,\sigma} 
\label{eq1}
\eear

The first two terms describe the c-orbital energy ($\ep_c$) and the kinetic energy of 
the conduction band arising from nearest neighbour hopping $t$, which is 
scaled as $t \propto \frac{t^*}{\sqrt{Z_c}}$ ($t^*$ is the unit of energy) in
the large dimension limit where the coordination number $Z_c \to \infty$. 
We choose the hyper cubic lattice for our calculation, for which 
the non-interacting density of states given by 
$\rho_0(\epsilon)=\exp{(-\epsilon^2/t_*^2)}/t_*\sqrt{\pi}$ is an unbounded 
Gaussian~\cite{geor}. The hybridization between the $c$- and $f$-electrons is 
represented by the third term ($V$) and is responsible for making otherwise 
localised $f$-electrons itinerant. The last term is $f$-orbital site energy ($\ep_f$)
and the on-site Coulomb repulsion $U$ for two $f$-electrons of opposite spin. 

In the non-interacting ($U=0$) case, the two orbital energies 
$\epsilon_c$ and $\epsilon_f$
completely determine the nature of the ground state, i.e.\ whether the system
would be gapped or gapless at the Fermi level. Equivalently, the occupation
numbers $n_f=\sum_\sigma \langle f_{i\sigma}^\dag
f^{\phantom{\dag}}_{i\sigma}\rangle $ and 
$n_c=\sum_\sigma \langle c^\dag_{i\sigma}c^{\phantom{\dag}}_{i\sigma}\rangle$ may also be used to characterize the 
ground state.
It is straightforward to see that the system is a band insulator for
$n_f+n_c=2$, while it is metallic for $n_f+n_c\neq 2$~\cite{aepp}. 
In the absence of hybridization, i.e.\ $V=0$, the $f$ and $c$ sites decouple
from each other. Since the $f$-part is site-diagonal in this limit,
the $f$-Green functions are trivial to obtain, e.g.\ through the equation
of motion method~\cite{econ}.  The tight-binding part along with the on-site
$c$-orbital energy may be solved through Feenberg renormalized
perturbation theory (FRPT)~\cite{econ}, and the non-interacting Green's 
function may be obtained as $g^c_0(\om)=(\om^+ - \ep_c - S(\om))^{-1}$.
The Feenberg self-energy $S(\om)$ is in fact a lattice-specific functional of
$g^c_0$; For example, $S[g^c_0]=\case{1}{4}t_*^2g^c_0$ for the Bethe lattice
in the limit of infinite dimensions~\cite{econ}.

Within DMFT, the $f$-selfenergy, $\Sigma(\om)$ 
is local~\cite{geor,prus95} in real space, hence the FRPT may be used 
for $V\neq 0$ also to obtain the $f$ and $c$-Green's functions. These are
then given by~\cite{vidh03,vidh04}
\bear
G^f(\omega)=\left[\om^+ - \ep_f - \Sigma(\om) - \frac{V^2}{\om^+ -
\ep_c - S(\om)}\right]^{-1}\label{eq2} \\
G^c(\omega)=\left[\om^+ - \ep_c - S(\om) - \frac{V^2}{\om^+ -
\ep_f - \Sigma(\om)} \right]^{-1}\label{eq3}
\eear
where $\Sigma(\om)$ is the momentum-independent $f$-self-energy 
representing interaction effects and $S(\om)$ is the same functional
of $G^c$ as it is of $g^c_0$ in the non-interacting limit.
This Feenberg self-energy may be obtained through the relations given below
(please see references~\cite{vidh03,vidh04} for more details):
\beq
G^c(\om)= H[\gamma] = \int^\infty_{-\infty} \frac{\rho_0(\ep)}{
\gamma(\om)-\ep} = \frac{1}{\gamma(\om)-S(\om)}
\label{eq:hilb}
\eeq
where $H[\gamma]$ denotes the Hilbert transform of $\gamma$ with respect
to the non-interacting density of states $\rho_0(\ep)$. While $\gamma=\om^+$
in the $V=0$ limit, for finite $V$, it is given by $\gamma(\om) = 
\om^+ - \ep_c - V^2[\om^+ - \ep_f-\Sigma(\om)]^{-1}$.

\subsection{Local moment approach: theory and practice}
\label{subsec:lma}
As mentioned in the introduction, within DMFT, the lattice model is 
mapped onto an impurity model with a self-consistently determined 
hybridization. And the solution of the resulting impurity problem
is the most challenging step in the solution of the lattice model.
We choose the local moment approach within DMFT to solve the PAM.
The LMA has been benchmarked extensively in case of the SIAM with other
exact methods such as NRG and Bethe Ansatz.
While in case of the PAM, excellent agreement with experiments for a 
wide variety of materials ranging from Kondo insulators to
heavy fermion metals and for a range of experiments such as dc and optical
transport as well as magnetotransport justifies its use to understand the
crossover from Kondo lattice regime to the empty orbital regime via the
mixed valence regime. The theoretical formalism and 
practical implementation are described in references ~\cite{vidh03,vidh04,vidh05}, and the reader is referred to these works for details.
\subsection{Transport: linear response}
Within DMFT, vertex
corrections are absent, hence the single-particle Green's functions
are sufficient within Kubo formalism to obtain transport quantities such as
dc resistivity and optical conductivity. The expressions derived
previously~\cite{vidh05} can be brought into a simpler form (derived in 
the appendix) given below:
\beq
\hspace{-3cm}
\sigma(\om)=\frac{\sigma_0}{2\pi^2} \Re \int^\infty_{-\infty} \,d\omp
\frac{n_F(\omp) - n_F(\om + \omp)}{\om} 
 \left[\frac{G^{c*}(\omp) - G^c(\om + \omp)}{\gamma(\om + \omp) - 
\gamma^*(\omp)} -\frac{G^c(\omp) - G^c(\om + \omp)}{\gamma(\om + \omp) - 
\gamma(\omp)}\right]
\label{eq4}
\eeq
where $\sigma_0=4\pi e^2 t^2 a^2 n/\hbar$ for lattice constant $a$,
electronic charge $e$, and electron density $n$ and $\gamma(\om)=
\om^+ - \ep_c - V^2[\om^+ - \ep_f - \Sigma(\om)]^{-1}$.
The DC conductivity is obtained by considering the $\om\rightarrow 0$
limit (see appendix for more details) of the above equation as
\beq
\sigma_{{\scriptstyle{DC}}} =  \frac{\sigma_0}{2\pi^2} \Re
\int^\infty_{-\infty} d\om\,\left(-\frac{dn_F}{d\om}\right) 
 \left[
\frac{\pi D^c(\om)}{\Im\gamma(\om)} + 
 2\left(1-\gamma(\om)G^c(\om)\right)\right]
\label{eq5} 
\eeq
where $D^c(\om)=-\Im G^c(\om)/\pi$ is the spectral
function of the retarded Green's function $G^c(\om)$ and $n_F(\om)=
(e^{\beta\om}+1)^{-1}$ is the Fermi function.

\section{Results}
In this section, we show results for spectral functions, i.e the density
of states (dos), at both zero and finite temperature, optical conductivity
again for $T\geq 0$, and DC resistivity. We work with a reasonably large interaction
strength ($U\simeq 4.5t_*$) and a hybridization strength of $V^2=0.6t_*^2$. The crossover
from Kondo lattice regime to the mixed-valent regime is investigated by
varying $\ep_c$ and $\ep_f$ such that the total occupancy $n_{tot}=n_c + n_f$ is fixed,
amd $n_f$ is decreased. This is done because pressure experiments would be expected
to keep the total occupancy fixed. The practical details of implementation are given
in references~\cite{vidh04,vidh05}. We begin with the low energy scale and the spectra.

\label{sec:res}

\subsection{Coherence scale}
\begin{figure}[h]
\centering{
\includegraphics[scale=0.4,clip=]{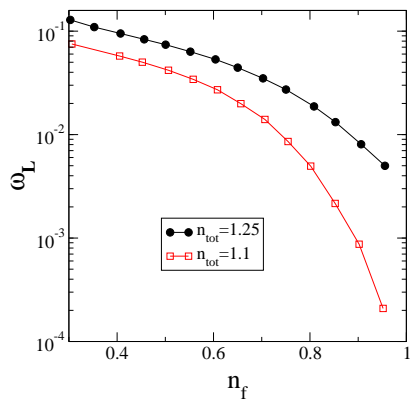}
}
\caption{A sharp decrease of the lattice coherence scale,
$\om_L$ is seen, with $n_f$ increasing from the empty orbital regime 
($n_f\ll 1$) through the mixed-valence regime
($0.3 \lesssim n_f \lesssim 0.8$) to the Kondo lattice 
regime ($n_f\rightarrow 1$). 
}
\label{fig:wl}
\end{figure}

The low energy coherence scale has been identified in previous studies
to be $\om_L=ZV^2/t_*$, where $Z=(1-\partial\Sigma/\partial
\omega|_{\om=0})^{-1}$ is the quasiparticle weight (inverse
effective mass). This scale is exponentially small in the strong coupling
Kondo lattice regime ($n_f\rightarrow 1$). Spectral functions, optical conductivities
and resistivity were shown to be universal functions of $(T/\om_L, \om/\om_L)$ in our
previous studies~\cite{vidh03,vidh04,vidh05}. 
Figure ~\ref{fig:wl} shows the variation of the lattice coherence 
scale $\om_L$ with the $f$-orbital occupancy $n_f$ for fixed 
$n_{tot}=1.25$ (filled circles) and $1.1$ (squares). 
A crossover from the Kondo lattice regime ($n_f\rightarrow 1$) through the 
mixed-valence regime ($0.3 \lesssim n_f \lesssim 0.8$) to the empty orbital
regime($n_f \ll 1$) manifests in a rapid increase in $\om_L$.
A lowering of the $f$-orbital occupation 
implies greater local charge fluctuations, implying an effective decrease
of correlation effects. This in turn implies a decrease in effective mass
or an increase in $Z$ and hence an increase in $\om_L$ with decreasing 
$n_f$.

\subsection{Density of states: Zero temperature}
\label{subsec:dos}

The density of states or the spectral functions are considered next. 
Changes in the low frequency Kondo resonance and in the high frequency 
Hubbard bands are identified.  

\begin{figure}[h]
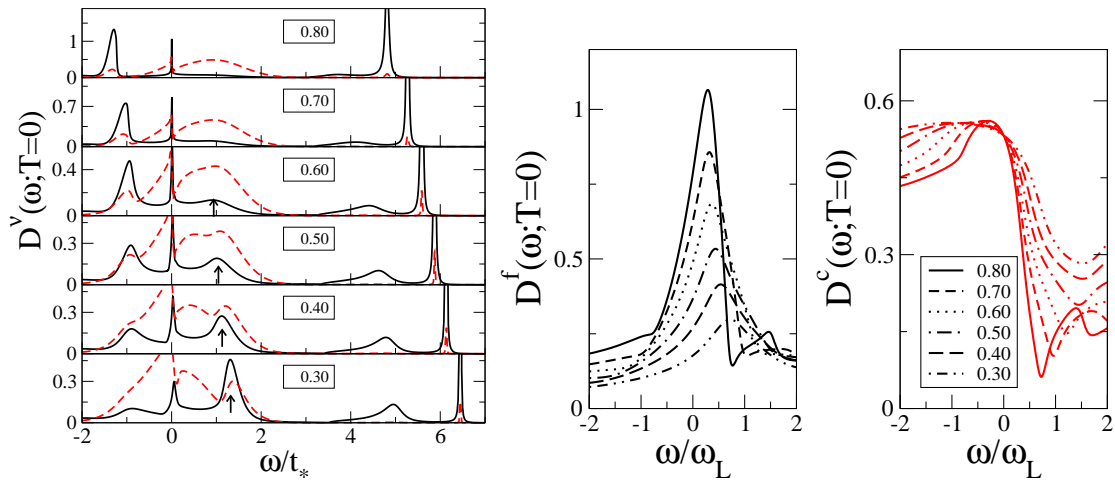

\centering{
\includegraphics[scale=0.5,clip=]{EPS/fig2_dos_allscales.eps}
\includegraphics[scale=0.6,clip=]{EPS/fig2_dos_lowf.eps}
}
\caption{(colour online) Left panel: The zero temperature spectral functions, 
$D^\nu(\om;T=0),\,\nu=c ({\rm dashed}),f ({\rm solid})$ are
 shown as a function of `absolute'
frequency $\om/t_*$ for various $f$-occupancies ($n_f$, indicated in the boxes and in the legends)
and a fixed $n_{tot}=1.25$.
The middle and the right panels show the same spectra as 
the left panel, but as a function of the scaled frequency
$\om/\om_L$ highlighting the changes in the Kondo resonance and the pseudogap.}
\label{fig:dos}
\end{figure}

We show the $T=0$ spectral functions ($D^\nu(\om;T=0)=-\Im G^\nu(\om,T=0)/\pi\;\nu=f,c$) 
as a function of the absolute frequency $\om/t_*$ in the left panel of figure~\ref{fig:dos}. 
In the $f$-dos (solid line), a usual three peak structure with high energy
Hubbard bands and a narrow Kondo resonance at the Fermi level is seen
for higher $n_f (\gtrsim 0.7)$. For lower $n_f$, a new spectral feature (marked by 
an arrow) appears above the Fermi level, the weight of which grows
with decreasing $n_f$. The lower Hubbard band (LHB) moves closer to the Fermi
level, while the upper Hubbard band (UHB) shifts to higher energies
with decreasing $n_f$.  The middle panel in figure~\ref{fig:dos} shows the 
Kondo resonance in greater
detail as a function of the scaled frequency $\om/\om_L$. 
Close to the Kondo lattice regime, a pseudogap is seen straddling the resonance,
which gets progressively filled up and for the lower values
of $n_f$, there is no trace of a pseudogap. Such a transfer of spectral
weight is very non-trivial and is a very
significant feature since it manifests clearly in a second peak in optical
conductivity, as will be discussed in sections~\ref{subsec:opt} and 
~\ref{subsec:inad}. 

The conduction band density of states, $D^c(\om;T=0)$, shown 
in the left panel (red dashed lines) of figure~\ref{fig:dos},
has an overall Gaussian envelope, with spectral weight carved out at the 
 effective $f$-level, $\ep_f^*$ (seen clearly  in the right panel of figure~\ref{fig:dos}
 and further discussed in section~\ref{subsec:ddos}) in the form
of a pseudogap and small Hubbard bands flanking the envelope.
In contrast to the $f$-dos (left panel of figure~\ref{fig:dos}), 
the Hubbard bands here
possess a very small fraction of the total spectral weight. The lower Hubbard
band is distinct from the envelope at higher $n_f (\gtrsim 0.6)$, and 
merges into the envelope at lower $n_f (\lesssim 0.5)$. The pseudogap
fills up with decreasing $n_f$ and for the lowest $n_f$ values shown,
there is indeed no trace of a pseudogap (see inset). The position of the Kondo resonance, the
Hubbard bands and a few other aspects are discussed in greater detail in subsection~\ref{subsec:ddos}.

\subsection{Density of states: Temperature dependence}
\label{subsec:tdos}

\begin{figure}[ht]
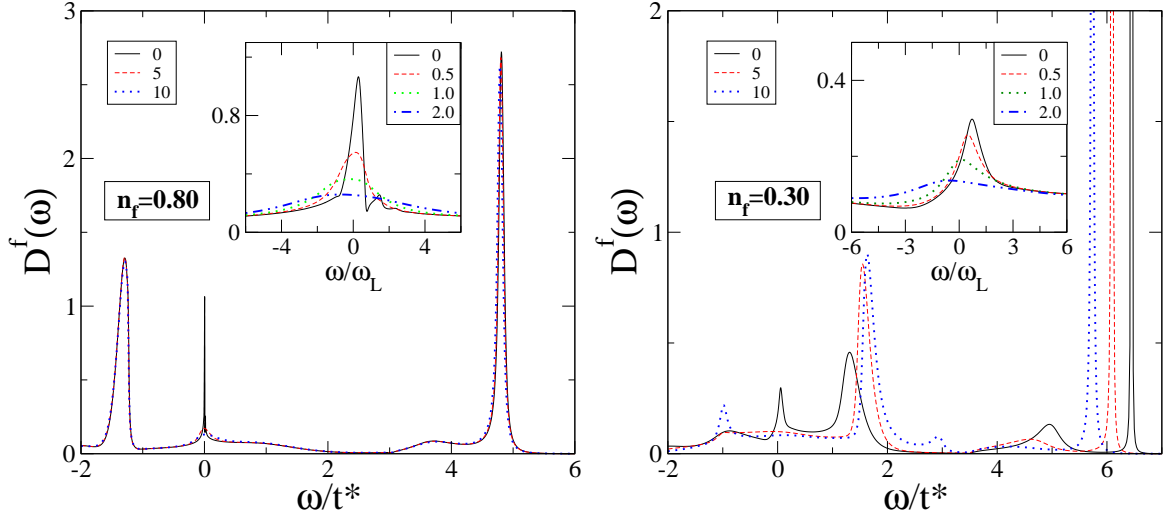

\centering{
\includegraphics[scale=0.55,clip=]{EPS/fig3_nf0.8_FT.eps}
\includegraphics[scale=0.55,clip=]{EPS/fig3_nf0.3_FT.eps}
}
\caption{
The $f$-electron spectral function at temperatures
$\tilde{T}=T/\om_L=0,2,5,10,20$ (indicated in legends) and for all scales 
in the main panel.
The insets show the low frequency part as a function of scaled frequency
for lower temperatures.
The left panel is for $n_f=0.80$ and the right panel is for $n_f=0.30$
(with fixed $n_{tot}=1.25$).
}
\label{fig:fsp_FT}
\end{figure}

In figure~\ref{fig:fsp_FT}, we show the evolution
of the $f$-dos as a function of temperature for $n_f=0.8$ and $n_f=0.3$
in the left and right panels respectively. The insets in the panels show
the low frequency, low temperature 
($(\om/\om_L,T/\om_L) \sim {\cal{O}}(1)$) part more clearly. 

The left inset shows that the pseudogap proximal to the Fermi level 
fills up, while a thermal broadening of the Kondo resonance is clearly 
visible in the $f$-dos.  At any given temperature $T$, it is expected
that changes in the spectral function will occur in the frequency range
$\om\lesssim T$. And indeed, this is seen in the KL regime as discussed below
and in reference ~\cite{vidh05}. 
The main left panel shows spectra for three 
different temperatures, which are hard to distinguish. This implies that
there has hardly been any spectral weight transfer on a scale of $t_*$,
even at the highest temperature shown ($\tilde{T}=10$, equivalent to
$\sim 0.18t_*$). 

The physics in the mixed-valent regime is, naturally, different.
The coherence scale is not exponentially small (see figure~\ref{fig:wl}).
In this regime, spectral weight transfer occurs
in an energy interval that is far greater than the thermal energy scale.
This may be seen in the right panel of figure~\ref{fig:fsp_FT}, 
where at a temperature of $5\om_L (\sim
0.77t_*)$, the UHB which is at $\sim 6t_*$ and hence about 15 times 
the temperature
gets strongly affected. In fact, the integrated spectral weight in the
UHB is found to decrease with $T$. Such differences in terms of 
spectral weight transfer 
between the KL and MV regime have been noted previously~\cite{vidh05}.
The new spectral feature above the Fermi level (marked by an arrow in the left panel of
figure~\ref{fig:dos}) that emerges distinctly in the MV regime
gains spectral weight and grows with increasing $T$. The reason for the growth
of this feature is not completely clear; however, since the total spectral weight is conserved,
the melting of Kondo resonance and the loss of spectral weight in the UHB
with increasing $T$ must be compensated
by gain in spectral weight elsewhere, and this could be one of the reasons for
the growth of this feature with increasing $T$. 

Angle-resolved photoemission experiments should be able to easily
identify such a feature. The conduction band density of states exhibits
similar temperature dependence as the $f$-dos and is hence not shown.

\subsection{$T=0$ optical conductivity}
\label{subsec:opt}

\begin{figure}[h]
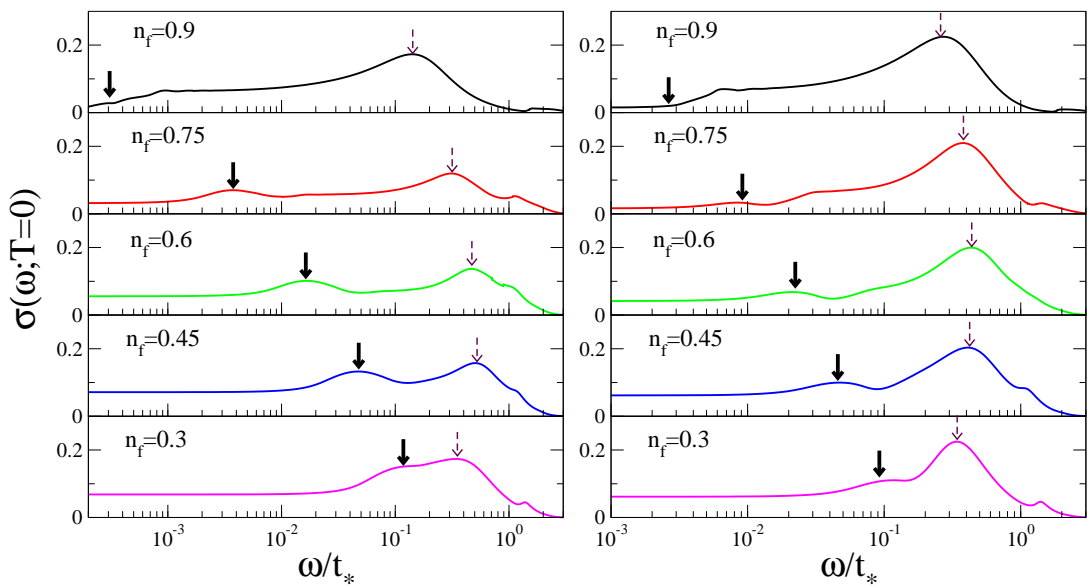

\centering{
\includegraphics[scale=0.55,clip=]{EPS/fig4a_opt_Teq0.eps}
\includegraphics[scale=0.55,clip=]{EPS/fig4b_opt_Teq0.eps}
}
\caption{The zero temperature optical conductivity $\sigma(\om;T=0)$ 
as a function of frequency ($\om/t_*$) for various $n_f$, and 
fixed $n_{tot}=1.1$ (left panel) and $1.25$ (right panel). 
The MIR peak is marked with a thin dashed arrow, while 
the DMIR feature is marked with a thick solid arrow.
}
\label{fig:opt}
\end{figure}

 As mentioned before, 
the zero temperature Green's functions are sufficient within DMFT to compute
the $T=0$ optical conductivity, $\sigma(\om;T=0)$ 
(equations~\ref{eq4})~\cite{geor,prus95}. 
In figure~\ref{fig:opt}, we show $\sigma(\om;T=0)$ as a function of frequency
($\om/t_*$) for various values of $n_f$ and a fixed $n_{tot}=1.1$. 
For $n_f \rightarrow 1$ , a familiar structure of
a single mid-infrared peak (MIR) arising due to interband 
transitions~\cite{okam07}, marked with a thin dashed arrow in 
figure~\ref{fig:opt}, is obtained. As we decrease $n_f$,
the MIR peak shifts to higher frequencies.
A Drude peak is obtained at the lower frequencies for all $n_f$, but it cannot 
be shown here since it is a Dirac delta function for $T=0$.

The lower $n_f$ values hold a surprise.
A deep mid-infrared (DMIR) absorption feature or a shoulder, marked with 
a thick solid arrow in figure~\ref{fig:opt}, emerges for
lower $n_f$ values. This is distinct from the above
mentioned MIR peak. Such a two peak structure 
has been reported earlier for the classic mixed valent compound CePd$_3$~\cite{buch}
and recently for YbIr$_2$Si$_2$~\cite{taku}. For the former compound, spin-orbit
splitting was argued to be responsible for the two peaks, but the latter has not been understood
quantitatively. Recently it has been pointed out through band structure
studies that the DMIR feature could be due to transitions to the 
effective $f$-level~\cite{kimu}. Here we confirm 
 that this feature could indeed arise from absorption
into the effective $f$-level at $\ep_f^*=Z(\ep_f +\Sigma(0))$, at which
frequency, a pseudogap appears in the spectral functions 
(figure~\ref{fig:dos}). Thus the scenario proposed here could be
relevant for both CePd$_3$ and YbIr$_2$Si$_2$.
A detailed discussion of the effective $f$-level, the pseudogap  and
transfer of spectral weight into the pseudogap region is given in
section ~\ref{subsec:inad}.

Here we must add that a two-peak structure is indeed obtained
for higher $n_{tot}$, such as $n_{tot}=1.25$ shown in the right panel of
figure~\ref{fig:opt} and hence is a generic feature in the crossover regime.
Furthermore, with decreasing $n_f$, the DMIR feature merges into the MIR peak,
without being able to develop into a full distinct peak for lower $n_f$, 
and hence gets harder to distinguish for higher $n_{tot}$.

We now move on to finite temperature transport.

\subsection{DC Resistivity as a function of temperature}

\begin{figure}[ht]
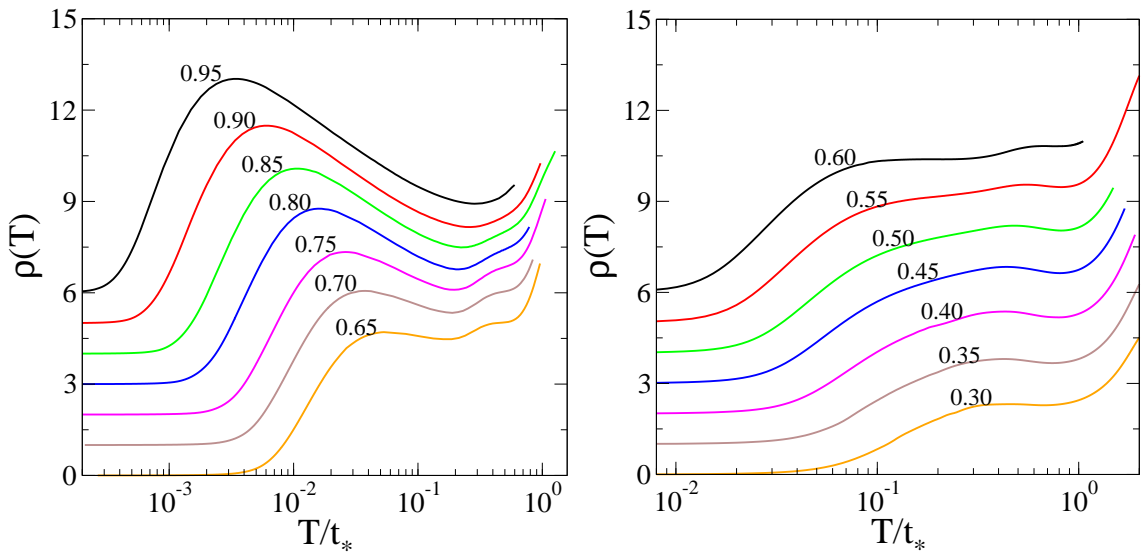

\centering{
\includegraphics[scale=0.59,clip=]{EPS/fig5a_dcres.eps}
\includegraphics[scale=0.55,clip=]{EPS/fig5b_dcres.eps}
}
\caption{DC resistivity for various $n_f$ values (mentioned next to the curve)
as a function of temperature $T/t_*$ for a fixed $n_{tot}=1.25$. 
 The theoretically computed $\rho(T)$ all have
zero residual values, but for visual clarity, they have been 
appropriately vertically offset. For example - in the left panel, 
we add a temperature independent $\rho_0$
that is zero for $n_f=0.65$ and increases by unity for
 every higher $n_f$ ($\rho_0=1,2,3...$ for $n_f=0.7, 0.75, 0.8,...$).
}
\label{fig:dcres}
\end{figure}

In figure~\ref{fig:dcres}, we show the dc resistivity as a function
of temperature $T/t_*$. 
The left panel shows the resistivity
for $0.65\leq n_f \leq 0.95$, while the right panel is for
$0.30\leq n_f \leq 0.60$. The theoretically computed $\rho(T)$ all have
zero residual values, but for visual clarity, they have been 
appropriately vertically offset (mentioned in the caption). There is a clear
shift of the coherence peak (low temperature peak) to higher temperatures
since its position correlates with the low energy scale~\cite{vidh05}, and
$\om_L$ itself increases sharply with decreasing $n_f$. 
For $n_f\lesssim 0.8$, a second maximum begins to form at higher temperature.
This second peak become easily distinguishable in the mixed-valence regime
and merges into the coherence peak to yield a broad maximum close to
the empty orbital regime (see right panel, $n_f\lesssim 0.45$).
Here we have shown that the crossover from the Kondo lattice regime to 
the mixed-valent regime involves the coalescing of the coherence peak and a high
temperature peak into a single broad peak. This high temperature
peak is not present in the KL regime or the empty orbital regime, and appears
in the mixed-valent regime only due to the proximity of the Hubbard band
to the Fermi level (discussed in detail in section~\ref{subsec:twopk}). 
Although we have displayed data in figure~\ref{fig:dcres} for $n_{tot}=1.25$,
the two peak structure is a generic feature of the mixed-valent regime 
for other $n_{tot}$ as well. However, the degree to which the two peaks 
can be distinguished and identified varies in different parameter regimes.

Pressure/doping-dependent
resistivity measurements in CeCu$_2$Si$_2$~\cite{jacc,ocko,fran01} show 
similar behaviour.
For low pressure, a two peak resistivity is observed. As a function
of increasing pressure, the two peaks merge and a single broad peak 
is seen. In the experimental paper, the effect of pressure was ascribed
to a crossover from Kondo lattice to intermediate valence, however 
the second peak was argued to arise from higher lying crystal-field
multiplets. Here, although we do concur with the crossover phenomenon, 
we present an alternative scenario for a two peak resistivity
structure. The scenario is that with increasing pressure, the valence 
decreases, bringing the Hubbard band in the proximity of the Fermi level,
thus leading to a second peak arising due to the Hubbard band states
contribution.  We emphasise here that a two-peak structure is obtained 
in our theory despite the model 
Hamiltonian (equation~\ref{eq1}) not containing crystal-field effects.  

We now move on to 
finite temperature optical transport.

\subsection{$T>0$ Optical transport}

\begin{figure}[ht]
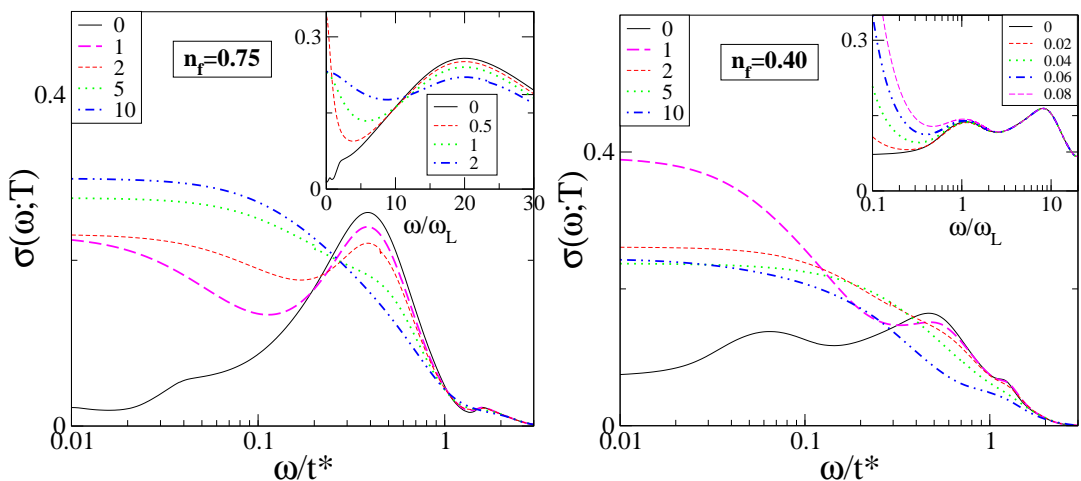

\centering{
\includegraphics[scale=0.55,clip=]{EPS/fig11a_opt.eps}
\includegraphics[scale=0.51,clip=]{EPS/fig11b_opt.eps}
}
\caption{Optical conductivity $\sigma(\om;T)$ for $n_f=0.75, n_c=0.61$ and 
$n_f=0.40, n_c=0.71$ in the left
and right panels respectively ($\ep_c=0.5t_*$ and $U\sim 4.7t_*$) 
as a function of $\om/t_*$ in the main panels
and as a function of $\om/\om_L$ in the insets respectively for
various temperature values (given as multiples of $\om_L$ in the legends).
The insets show $\sigma(\om;T)$ for low temperatures to show the
transfer of spectral weight more clearly.}
\label{fig:opt_FT}
\end{figure}

We now show the temperature dependence of the optical conductivity in 
figure~\ref{fig:opt_FT} for two different $f$-occupancies, $0.75$ and $0.4$ (other parameters
are mentioned in the figure caption) in
the left and right panels respectively. The $\sigma(\om;T)$ for the higher
 $n_f$ case has a usual prominent mid-infrared (MIR) peak, with a slight 
shoulder at lower frequencies. The inset shows the low temperature evolution
 of $\sigma(\om)$, while the main panel shows $T\leq 10\om_L$. With increasing
temperature, a transfer of spectral weight over all energy scales in seen.
The lower $n_f (0.4)$ optical conductivity shown in the right panel of figure
~\ref{fig:opt_FT} is qualitatively different from that of the left panel.
The shoulder like structure seen in the Kondo lattice regime (left panel)
develops into a full peak in the mixed-valent regime, while the MIR
peak diminishes in intensity and gets broader. The analysis at $T=0$ shows (see section~\ref{sec:disc})
that the lower frequency peak occurs at the effective $f$-level $\ep_f^*$.
The inset shows that the DMIR peak or the shoulder appears only below a 
temperature $T=0.08\om_L$ even though it is positioned at a frequency
$\om$ equal to the coherence scale $\om_L$ (since here $\ep_f + \Sigma(0)\sim 1$).
This precise behaviour is seen in recent optical conductivity measurements
in YbIr$_2$Si$_2$~\cite{taku}(see section~\ref{sec:comp}).

\section{Discussion}
\label{sec:disc}

\subsection{Spectral features: Kondo resonance and Hubbard bands}
\label{subsec:ddos}

In this subsection, we discuss the positions and weights of the various features seen in the spectra
in section~\ref{subsec:dos}. At high energies, the prominent spectral features are the Hubbard bands. 
In figure~\ref{fig:dos}, it is seen that the Hubbard bands
shift to the right with decreasing $n_f$. This is
 expected from the atomic limit,
since in that limit, the Hubbard bands occur at $\ep_f$ and $\ep_f+U$, and
$\ep_f$ moves closer to the Fermi level when $n_f$
decreases. However, we find from our calculations, that in the displayed spectra, the Hubbard
bands are not at the positions predicted by the atomic
limit, as must be expected, due to  a  combination of shifts in the levels
arising through hybridization and self-energy effects. 
We see that even for $n_f\sim 0.8$, the LHB and the UHB are not symmetrically
placed about the Fermi level as would have been the case in the strong coupling KL regime.
In fact, the LHB is at $\om\lesssim -t_*$, while the UHB is at $\om\gtrsim U$.
The LHB is predicted very well by the unrestricted Hartree Fock (UHF) solution
(not discussed here) which considers the static part of the self energy and hybridization effects. 
For predicting the UHB, we can neglect hybridization effects, however, we need 
to retain the static part and {\em most importantly} the real part of the self energy.
The latter may be approximated by an expression obtained through 
high frequency moment expansion~\cite{kaju}, and thus the UHB turns out to be at
$\sim (\bar{\ep_f} + \sqrt{\bar{\ep_f}^2 + \bar{U}^2})/2$, where
$\bar{\ep_f}=\ep_f + U(\bar{n} + \bar{\mu})/2$,  $\bar{U}^2 = U^2\bar{n}(2-\bar{n})$
and $\bar{n},\bar{\mu}$ have been defined in section 4.2 of reference~\cite{vidh04}
as the fictitious occupation number
and the moment obtained from the spin-dependent host/medium propagators that are used
to construct the LMA self-energies.

In the other extreme, i.e close to the Fermi level,
a low frequency form of the Green's functions using a Fermi liquid self-energy 
$\Sigma(\omega)=\Sigma(0) +\omega(\frac{1}{Z}-1)+{\cal{O}}(\omega^2)$
may be derived that enables us to understand 
most of the features of the Kondo resonance. 
The resulting spectral functions in the neighbourhood
of the Fermi level are given by
\bear
D^c(\om) &\sim   \rho_0\left(\om - \ep_c - 
\frac{ZV^2}{\om - Z\bar{\ep}_f}
\right) \nnu \\
V^2D^f(\om) &\sim  \left(\frac{ZV^2}{\om-Z\bar{\ep}_f}\right)^2
 \,D^c(\om)
\label{eq6}
\eear
where ${\ep}_f^*=Z\bar{\ep}_f=Z(\ep_f+\Sigma(0))$ is the
effective $f$-level. These equations show that the low energy 
($\om\lesssim \om_L$) spectral features
are precisely those of a non-interacting ($U=0$) PAM with renormalized 
parameters ($V^2\rightarrow ZV^2, \ep_f\rightarrow Z(\ep_f+\Sigma(0)$)
and  ($\ep_c\rightarrow \ep_c$)~\cite{vidh04}. 
Thus, this is the renormalized non-interacting limit (RNIL). 
A scaling collapse of 
the numerically obtained spectra with the analytical expressions above would 
be a demonstration
of adiabatic continuity of the interacting system to the non-interacting limit.
Indeed, we do see such a collapse (not shown), implying adiabatic continuity, although the 
range of frequencies over which such a collapse occurs decreases from $\om \lesssim \om_L$
in the KL regime, to $\om \ll \om_L$ in the empty orbital regime.

As $\om\rightarrow \ep_f^*$ in equations~\ref{eq6}, the renormalized non-interacting
limit spectral functions also vanish,
thus the pseudogap seen in the spectra
is positioned at the effective $f$-level, $\ep_f^* = Z \left(\ep_f+
\Sigma(0)\right)$.  Transfer of spectral
weight into the pseudogap happens with decreasing $n_f$ as seen in figure~\ref{fig:dos}
which implies that the frequency interval in the neighbourhood of $\ep_f^*$ is gaining
spectral weight. This spectral weight transfer is completely missed by 
theories that are equivalent to the renormalized non-interacting limit.
This includes approaches such as slave-boson theories or two-band 
models of heavy fermion systems, which ignore the imaginary part of self-energy. 
The variation of $\ep_f^*$ with $n_f$ may be easily predicted using
the Luttinger's theorem which states~\cite{vidh04}
\beq
\frac{1}{2}(n_f+n_c) = \int_{-\infty}^{-\ep_c+V^2/\bar{\ep}_f}
\rho_0(\ep)\,d\ep  + \theta(-\bar{\ep}_f) \,.
\label{eq9}
\eeq
For fixed $n_{tot}=n_f+n_c$, the upper limit of integration on the 
right side of equation~\ref{eq9} is also fixed. Crossing over from
the KL to MV regime would require decreasing $n_f$ and increasing $n_c$,
which in turn would require decreasing $\ep_c$. Thus to keep
$-\ep_c+V^2/\bar{\ep}_f$ fixed, the $\bar{\ep}_f=\ep_f + \Sigma(0)$ must increase,
and hence the effective $f$-level, $\ep_f^*=Z\bar{\ep}_f$, also increases as $n_f$ decreases.
This is indeed seen in the figure~\ref{fig:dos} because the position of the pseudogap
does indeed shift to higher frequencies as $n_f$ is decreased.

\subsection{Inadequacy of the renormalized non-interacting limit}
\label{subsec:inad}

In ~\cite{vidh05}, it was shown that the two band
model or the RNIL predicts a square root singularity at the minimum direct gap
($\sim 2\sqrt{Z} V$), and hence the MIR peak is generally attributed 
to the direct gap. A simple Fermi-liquid analysis $\Sigma(\om) \simeq
\Re\Sigma(0) + \om(1-1/Z)$ of the poles of the 
$\k$-dependent conduction electron Green's function 
\beq
G^c(\om;\ep_\k)= \left[\om^+ - \ep_c- \ep_\k - \frac{V^2}{\om^+
-\ep_f - \Sigma(\om)}\right]
\label{eq10}
\eeq
yields a two-band model
\beq
\om_\pm(\ep_\k)= \frac{\left(\ep_c + \ep_\k - \ep_f^*\right) \pm 
\sqrt{\left(\ep_c + \ep_\k - \ep_f^*\right)^2 + 4ZV^2}}{2}
\label{eq11}
\eeq
The minimum direct gap is given by ${\rm min}(\om_+(\ep_\k) - \om_-(\ep_\k))$.
The square root singularity at the direct gap appears in the two band model 
because the imaginary
part of the self energy is neglected, and the resulting spectral functions
are Dirac delta functions. Including incoherent effects due to
electron-electron scattering results in broadening of the MIR peak and cutting
off the square root singularity. 

At the lowest $n_f$ values, a clear two-peak structure is visible in the optical
conductivity displayed in the figure~\ref{fig:opt} (especially for $n_f=0.45$
in the left panel). The low frequency peak is
 in fact at the effective $f$-level, as argued
below, while the high frequency peak is the usual MIR peak.
Naively, finding an absorption peak at $\ep_f^*$ is counter-intuitive, 
because the renormalized
non-interacting limit (equations ~\ref{eq6}) shows that a pseudogap
exists at $\ep_f^*$, implying that there is no density of states at that
energy. So how can absorption into a gap happen? The answer is of course
that the RNIL, which is equivalent to a slave-boson mean-field theory, 
which in turn is
equivalent to a two-band model~\cite{okam07}, are not totally correct in their predictions.
These approaches neglect the imaginary part of the
self-energy (scattering rate). So even though the RNIL predicts a gap,
there is in fact no gap at $\ep_f^*$ when self-energy effects are included
(see the pseudogap feature in the density of states in figure~\ref{fig:dos},
and the discussion in section~\ref{subsec:ddos}).
And in fact, it may be shown rigorously, without recourse to LMA
that absorption to $\ep_f^*$ will happen provided  the imaginary part
of self energy is non-zero at that energy. To see this, 
consider the roots of the real part of the denominator of the 
conduction electron Green function,
$G^c(\om,\ep_\k)$ (equation~\ref{eq10}), given by
\bear
 \Re\left[\om - \ep_c - \ep_\k - V^2\left(\om - \ep_f - \Sigma(\om)\right)^{-1}
\right]= \nnu \\
\om - \ep_c - V^2\frac{\om - \ep_f - \Re\Sigma(\om)}{
\left(\om - \ep_f - \Re\Sigma(\om)\right)^2 + \left(\Im\Sigma(\om)\right)^2} 
= \ep_\k
\label{eq12}
\eear
If the imaginary part of the self energy is completely neglected and 
a first order Taylor expansion is carried out for the real part of the self energy, we get
back equation~\ref{eq11}. However, retaining the imaginary part,
however small it might be, results in an equation that is at least 
cubic in order. And for $\ep_\k= \ep_c-Z(\ep_f +\Sigma(0))$, one of the roots
is just $\ep_f^*=Z(\ep_f +\Sigma(0))$. 

\begin{figure}[t]
\centering{
\includegraphics[scale=0.55,clip=]{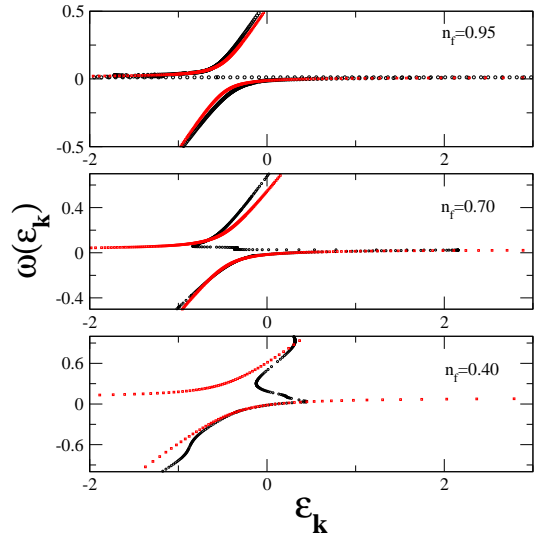}
}
\caption{The dispersion $\om(\ep_\k)$ computed from equation~\ref{eq12} (black)
and from the RNIL, equation~\ref{eq11}(red) for $\ep_c=0.5t_*, U\sim 4.7t_*$
and $n_f=0.95, n_c=0.58$ (top panel),
$n_f=0.70, n_c=0.63$  (middle panel ) and $n_f=0.40, n_c=0.71$ (bottom panel).
}
\label{fig:disp}
\end{figure}

   We support the arguments above with LMA results below.
The dispersion, $\om(\ep_\k)$ is computed for $n_f=0.95 (n_c=0.58), 0.7 (n_c
=0.63)$ and $n_f=0.4 (n_c=0.71)$ with $\ep_c=0.5, U\sim 4.7t_*$
through 
equation~\ref{eq12} and shown in the top, middle and bottom panels 
respectively in figure~\ref{fig:disp}. The two-bands obtained at the
renormalized non-interacting level (eq.~\ref{eq11}) are also 
superimposed in red.
It is clear from the figure that the agreement of the two-band model 
with the full dispersion gets progressively worse as $n_f$ decreases.
For $n_f=0.4$, the middle band is clearly visible in the full dispersion,
while being completely absent in the two-band picture. This third band, 
as argued above is centred at $\om=\ep_f^*$. It is
the excitations from the `band' below the Fermi level to $\ep_f^*$, that
appears as an additional `anomalous' absorption peak in the deep mid-infrared 
region (as a shoulder). This anomalous peak is seen to become prominent in the
mixed-valent
regime and is very small or invisible in the Kondo lattice regime.  We refer
the reader to figure~\ref{fig:opt}, where an extra absorption feature
is easily discernible for lower $n_f$ values, while for higher $n_f$
values, just a weak shoulder is observed. As we will see later,
such a two-peak optical conductivity has indeed been observed in recent experiments (section\ref{sec:comp}).

\subsection{Two-peak resistivity: why?}
\label{subsec:twopk}
We have shown previously that the coherence peak 
(low temperature peak) in the resistivity, which would be a minimum in 
the conductivity~\cite{vidh05},
occurs at a temperature comparable to $\om_L$, the low temperature scale.
A second peak (high temperature peak) has indeed been observed in experiments
and has been attributed to crystal field split levels(for example in 
Ref~\cite{jacc}).
Notwithstanding the foregoing possibility, we find an alternative and
 far simpler explanation for the existence of the second peak 
even within the single
band PAM. The second peak occurs at a temperature that is roughly
half of the lower Hubbard band energy scale. It might at first seem surprising
to note that the Hubbard band is contributing to transport. 
The Hubbard bands are usually at an energy scale of $U/2$ in the strong
coupling Kondo lattice regime which being
of the order of a few eV, remain untouched until room temperature.
Nevertheless, for mixed-valent systems, even if the $U$ is large, either the LHB
or the UHB moves close enough to the Fermi level so as to be affected
at room temperature scales. 
\begin{figure}[ht]
\centering{
\includegraphics[scale=0.45,clip=]{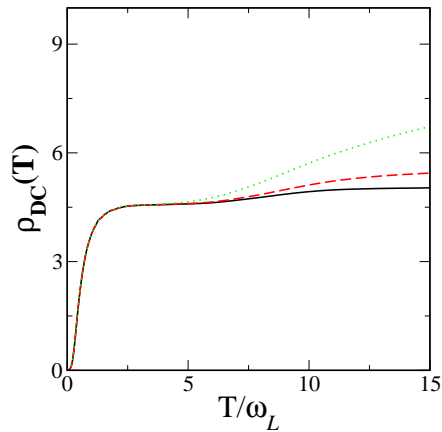}
}
\caption{Figure shows that the second peak in the DC resistivity seen
most clearly in the right panel of figure~\ref{fig:dcres} for $n_f=0.6$
disappears progressively by excluding the lower hubbard band and other
spectral features.  The solid line is for $n_f=0.6, n_c=0.66$ and is the
full resistivity. The dashed and the dotted
lines are obtained by excluding the lower Hubbard band below the
Fermi level and with decreasing cutoff above the Fermi level 
(see text for discussion).
}
\label{fig:hubpk}
\end{figure}

To show the contribution of the Hubbard band states to the conductivity,
we consider expression~\ref{eq5} again. On the right hand side,
the integration is carried out over all frequencies. However, if we introduce
a upper and a lower cutoff in the integration limits, we can isolate the
contribution of individual spectral features within those limits. 
In figure~\ref{fig:hubpk},
we show the calculated dc resistivity for $n_f=0.6, n_c=0.65$ (same as 
that in figure~\ref{fig:dcres}), with three
 different cutoffs: (i)the solid line being no cutoff (full resistivity),
(ii) the dashed line having limits such that the lower Hubbard band is excluded
but the new `non-Hubbard band' feature above the Fermi level 
(marked with an arrow in 
figure~\ref{fig:dos}) is included and
(iii) the dotted line having limits such that both the LHB and the new feature 
is excluded. It is seen that excluding the spectral weight of the 
Hubbard band and the new feature enhances the resistivity systematically.
This shows that the Hubbard band contribution to the dc conductivity is 
substantial in mixed-valent compounds.
The two-peak behaviour of resistivity, is not specific to the parameter 
regime for which the dc resistivity has been displayed 
($U/V^2\sim 8, n_{tot}=1.25$).  In other parameter 
regimes, such as $n_{tot}=1.1$, our calculations demonstrate 
(not shown here) that the
dc resistivity crosses over from a single sharp coherence peak to a 
broad peak lineshape with decreasing valence, going via a 
two-peak structure but with very faint or hard to distinguish features. 

\section{Comparison to Experiments}
\label{sec:comp}

\subsection{DC resistivity: ambient pressure}
\begin{figure}[ht]
\centering{
\includegraphics[scale=0.61,clip=]{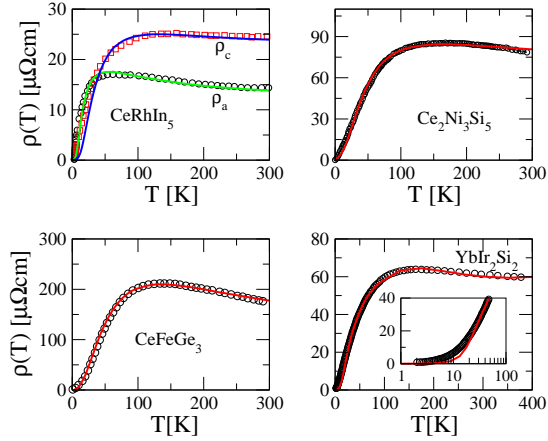}
}
\caption{Comparison of resistivity measurements for four different
materials with theory. Please refer to text for discussion. 
}
\label{fig:vardcres}
\end{figure}

In figure~\ref{fig:vardcres}, we have superposed theoretically computed dc resistivity
(solid lines) with the experiments (circles) for CeRhIn$_5$~\cite{chri}, Ce$_2$Ni$_3$Si$_5$~\cite{chan}, CeFeGe$_3$~\cite{budk}
and YbIr$_2$Si$_2$~\cite{taku}. To find the appropriate theoretical parameters for each material, we 
had to adopt a trial and error approach. However, a surprising similarity in the dc resistivity
of CeRhIn$_5$ and Ce$_2$Ni$_3$Si$_5$ reduced our effort substantially.
We found that we could take the experimentally measured dc resistivity for the two materials
and scale them onto each other simply by rescaling the $x$ and $y$ axes. CeRhIn$_5$ has anisotropic
resistivity, nevertheless the $\rho_a$ and $\rho_c$ may also be scaled onto each other, thus
showing that qualitatively, they are also similar. We find that these two materials have an
$f$-occupancy $n_f\sim 0.7$, with the rest of the parameters being $n_c\sim 0.55,  U/V^2\sim 8,
\ep_c=0.5t_*$. For CeFeGe$_3$, we found that the mixed-valence resistivities do not fit the data
well. Rather, the best fit was found using parameters ($n_f\sim 1, n_c\sim 0.77, U/V^2\sim 5, \ep_c=0.3$)
that signify an intermediate correlation with $f$-occupancy being nearly unity.
The optical transport data for YbIr$_2$Si$_2$ (see below) shows a two-peak structure very similar
to that seen in figure~\ref{fig:opt} for $n_f=0.4$ and $n_{tot}=1.1$. Thus we take the same resistivity
and superimpose that onto the experimentally measured one, and we see very good agreement. 

\begin{figure}[th]
\centering{
\includegraphics[scale=0.55,clip=]{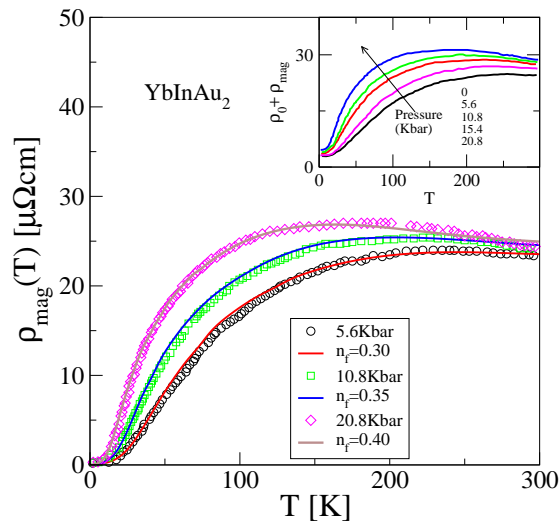}
}
\caption{Comparison of pressure dependent resistivity measurements (magnetic contribution
only) for YbInAu$_2$ with theory.
The inset shows the reported~\cite{fuse} experimental data.}
\label{fig:ybinau2}
\end{figure}
\subsection{DC resistivity: Pressure dependence}
Hydrostatic pressure dependence measurements of the dc resistivity of YbInAu$_2$ and its non-magnetic
homologue LuInAu$_2$ have been carried out by Fuse et.al~\cite{fuse}. The experimental magnetic resistivity along with
the residual resistivity ($\rho_{mag}+\rho_0$) is shown in the inset of figure ~\ref{fig:ybinau2}.
The main panel shows a comparison of theory (solid lines, $n_{tot}=1.1, U/V^2\sim 8$) 
with experiment (symbols). The agreement
is seen to be excellent. The $T_{max}$ is seen to decrease with pressure, and the material appears to be 
progressing towards a Yb$^{3+}$ state with increasing pressure, as conjectured in the experimental paper.
Nevertheless, the valence remains in the mixed-valent regime ($\lesssim 0.4$), even with pressures upto 20KBar. 

\begin{figure}[ht]
\centering{
\includegraphics[scale=0.23,clip=]{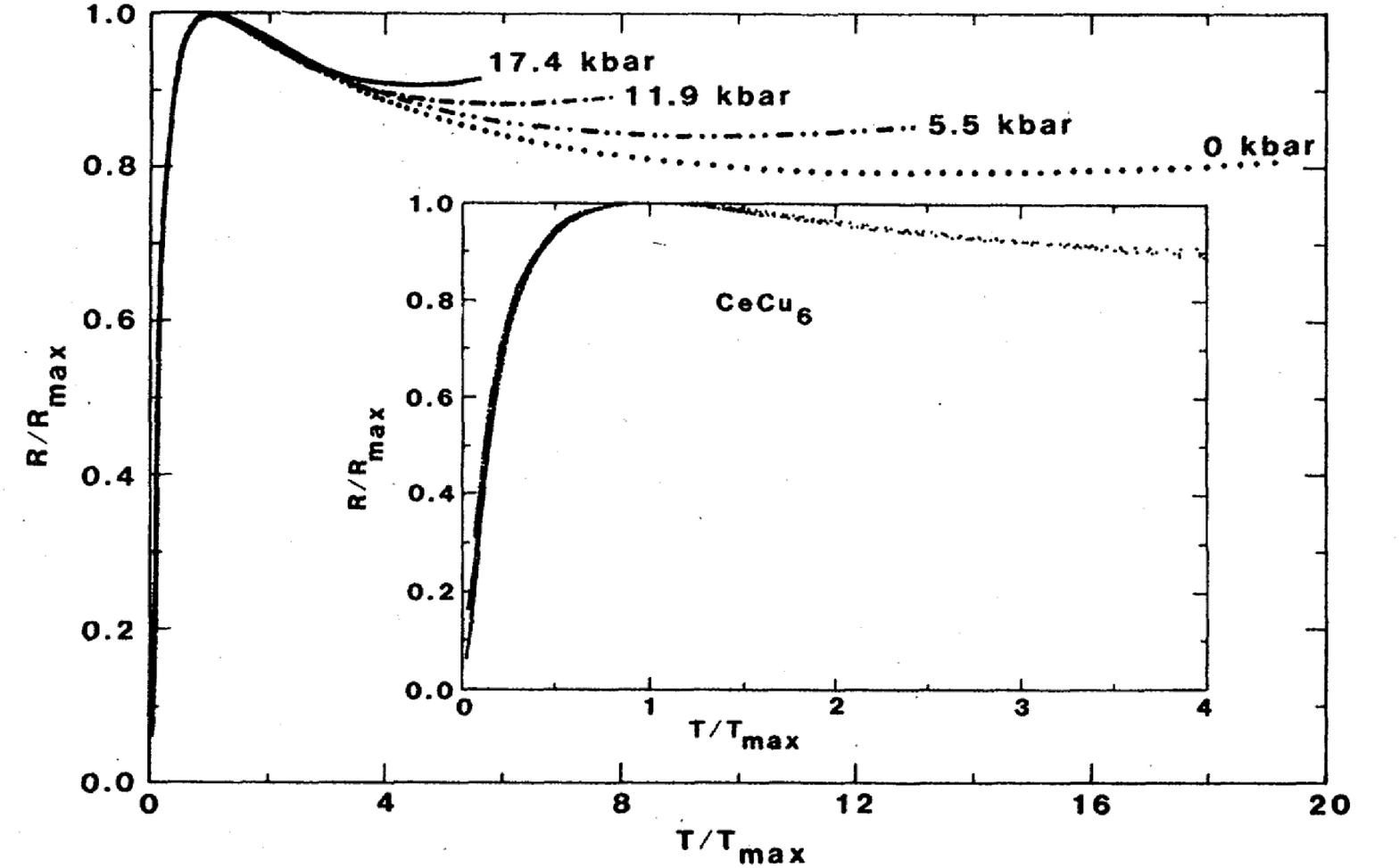}
\includegraphics[scale=0.51,clip=]{EPS/fig11_cecu6_th.eps}
}
\caption{Left panel: Experimentally measured pressure dependent resistivity measurements
for CeCu$_6$~\cite{thom}. Right panel: Theoretically computed $\rho(T)$ (same data as 
left panel of figure~\ref{fig:dcres}) plotted as $\rho/\rho_{max}$ {\it vs.} $T/T_{max}$.
The respective valencies are mentioned next to the curves.}
\label{fig:cecu6}
\end{figure}

We now move on to CeCu$_6$, which has gathered a lot of attention in the past decade
as a material that can be tuned to a quantum critical point with Au doping. Pressure
dependent resistivity measurements on CeCu$_6$ were carried out in 1985 by Thompson and Fisk~\cite{thom}.
They found that with increasing pressure, the material crosses over from Kondo lattice like
to mixed-valence like regime. In the left panel of figure~\ref{fig:cecu6}, we show the experimental
graph, while in the right panel, the theory (same data as that for figure~\ref{fig:dcres}) is shown.
The experiment shows that with increasing pressure, the temperature dependence follows
the $P=0$ curve to some temperature and then deviates. In the inset of the left panel,
the experimental data for $0\leq P\leq 17.4$kbar is shown to collapse when plotted as
$R/R_{max}$ {\it vs.} $T/T_{max}$ for $T\leq 4T_{max}$. The theoretical curves in the right panel
correspond very closely to those seen in the experiment.
The agreement between theory and experiment shows that indeed with increasing pressure, the occupancy does
change from $n_f\rightarrow 1$ to $n_f\sim 0.75$, thus implying that the valence of Ce changes from $\sim 3+$ to 
$\sim 2.75+$. The theory curves decrease more rapidly (than experiment) with increasing temperature 
beyond the coherence peak, and the reason for this is that the phonon contribution is not subtracted
in the experiment. The inset in the theory panel has five different valencies plotted together, namely,
$n_f=0.90, 0.85, 0.80, 0.78$ and $0.75$. Except for $n_f=0.75$, the rest of the data is seen to 
collapse onto a single curve, as seen in the experiment. The $n_f=0.75$ curve does collapse
upto a certain temperature and then deviates from the universal curve.

\subsection{Optical transport: YbIr$_2$Si$_2$}

\begin{figure}[ht]
\centering{
\includegraphics[scale=0.49,clip=]{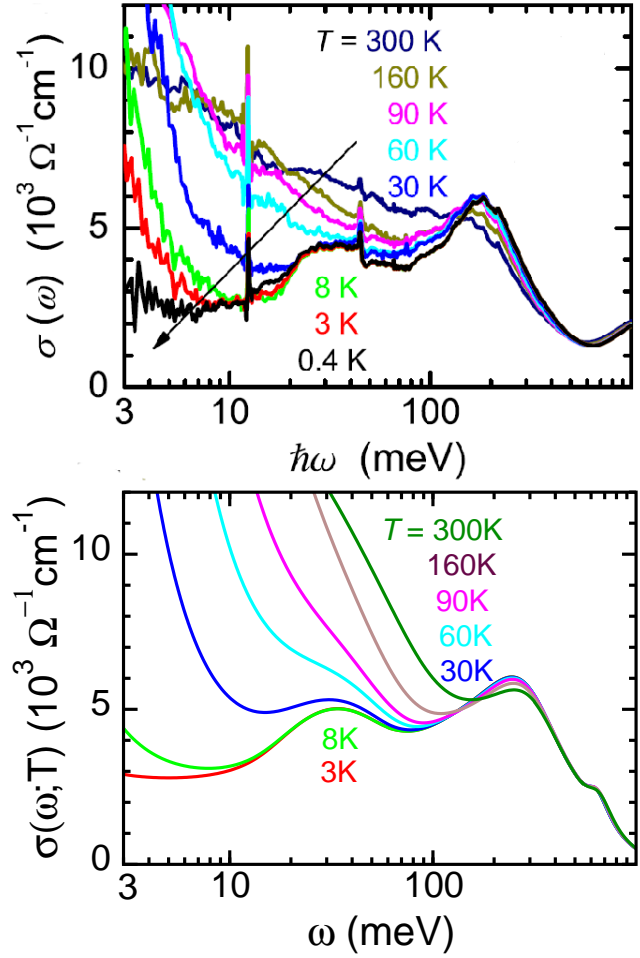}
\includegraphics[scale=0.61,clip=]{EPS/fig12_YbIr2_th.eps}
}
\caption{top panel: Experimentally measured optical conductivity 
for YbIr$_2$Si$_2$.~\cite{taku} Bottom panel: Theoretically computed $\sigma(\om;T)$
for $U/t_*\sim 5, V^2\sim 0.6t_*^2$ and $\ep_c=0.5t_*$, which yields
$n_f=0.4$ and $n_c=0.7$. Excellent
quantitative agreement is observed between theory and experiment for the 
line shape and the temperature dependence.}
\label{fig:opt_YbIr2}
\end{figure}

The recently discovered heavy fermion system YbIr$_2$Si$_2$~\cite{hoss} has a 
crystal structure similar to the well studied YbRh$_2$Si$_2$. The latter 
exhibits a field-tuned
quantum critical point (QCP), while the former has a pressure-tuned first
order phase transition to a ferromagnetic phase.
The experimentally measured optical conductivity~\cite{taku} is shown in the
top panel of figure~\ref{fig:opt_YbIr2}. A clear two peak structure
is evident, and a large scale spectral weight transfer occurs 
as temperature is increased from 0.4K to 300K. An important 
characteristic of the temperature dependence of $\sigma(\om;T)$ is
that, as temperature is increased, the lower frequency peak (shoulder) merges into the continuum
at $T\sim 60$K, at which temperature the higher frequency peak remains
untouched. 

We note that the theoretically computed optical conductivity shown 
in figure~\ref{fig:opt_FT} does appear to resemble the experimentally
measured $\sigma(\om;T)$.  In the experiment, the shoulder peak appears
 only for $T\lesssim 30K$,
which, according to theory, should be roughly $0.08$ times the peak 
frequency for the chosen parameters(see the discussion for 
figure~\ref{fig:opt_FT}). 
Thus, theoretically, we can predict that the shoulder should appear
at $\sim 30.0/0.08 K \simeq 32 meV$. The shoulder peak position as predicted
by theory indeed agrees very well with the experimentally observed shoulder
position ($\sim 30 meV$). The bottom panel of the same figure shows the 
optical conductivity
computed for $U/t_*\sim 5, V^2\sim 0.6t_*^2$ and $\ep_c=0.5t_*$, which yields
$n_f=0.4$ and $n_c=0.7$, thus classifying YbIr$_2$Si$_2$ as a mixed-valent
material. These parameters were chosen using the results
shown in figure~\ref{fig:opt}, and for consistency, we note that the 
dc resistivity in the experiment has a broad and shallow peak  
(see bottom right panel of figure~\ref{fig:vardcres}) which indicates
that YbIr$_2$Si$_2$ belongs to the mixed-valent regime.
The DC conductivity obtained through a low frequency
limit in the theory is  higher than the corresponding
experimental values. This is natural, since the theory neglects electron-phonon
scattering, which if included would reduce the theoretical conductivities.
Apart from this disagreement, an excellent agreement between theory and
experiment is seen in quantitative terms for both the lineshape and
the temperature dependence.
A phenomenological analysis ~\cite{taku} of the experimental optical conductivity
 using Drude and extended Drude
formalism has been used to infer  non-Fermi liquid nature of the 
quasiparticles in this material. However, if the quantitative agreement 
between theory and experiment is any indication, this material is 
a perfect Fermi liquid. The self energy has the correct Fermi liquid form 
for the parameters used to compute the optical conductivity in the 
bottom panel of figure~\ref{fig:opt_YbIr2}. In our approach, the Luttinger's 
theorem is used as a constraint, which also supports the inference of 
Fermi liquid behaviour. The low temperature resistivity also exhibits
a clean $T^2$ behaviour, both in the experiment~\cite{yuan} and 
theory (shown to agree
well in the inset of the bottom right panel in figure~\ref{fig:vardcres}),
 which again supports a Fermi liquid ground state.
Additionally, we conclude that
this material is in the mixed-valent regime with an effective $f$-occupancy
of $n_f=0.4$, which would correspond to an effective valence Yb$^{2.4+}$.
(Naturally, the $n_f$ given here must be interpreted as hole occupancy
for Yb compounds.)
This effective valence predicted by our theory disagrees with the 
inference from susceptibility measurements,
which yields a Yb valence close to $3+$. Further theoretical and experimental
investigations are needed to resolve this.
We can also infer that the effective $f$-level given by $\ep_f^*=Z(\ep_f
+\Sigma(0))$ in YbIr$_2$Si$_2$ is located at roughly 25-30 meV. 
The dc resistivity calculated from theory for the parameters mentioned
in the figure~\ref{fig:opt_YbIr2} does agree qualitatively with experiment
in terms of a broad peak and the lineshape(see figure~\ref{fig:vardcres}), 
however, the scale inferred from fitting to
the experimental data does not agree with that obtained from optical 
conductivity.  There appears to be an inconsistency in theory, in terms 
of an ambiguity in the parameters that can fit the experiment. 
The implication is, we believe, minor. 
It should be possible by a better scanning of the parameter space
to find a `right' set of parameters that can fit the optical and dc conductivity
 consistently with the same scale. We believe that finding this
right set of parameters will surely make the theory more convincing, but
probably not by a great measure, since almost all aspects of the experiment
are quantitatively captured by the presently used set of parameters. 
The new set of parameters would probably change the numbers (e.g for the
$U,V^2$ values), nevertheless, the major conclusions such as that of
YbIr$_2$Si$_2$ being a Fermi liquid and a mixed-valent system should 
remain unscathed.

\section{Conclusions}
\label{sec:conc}

We have carried out an extensive study of the changes in spectral and transport
properties of the periodic Anderson model as the $f$-occupation number is varied from unity to nearly zero. The local moment approach within dynamical mean 
field theory has been employed for this study. We have used the results
of our study to understand the crossover from mixed-valent regime to the 
Kondo lattice regime (or vice-versa) observed to occur in many rare earth 
intermetallics. We find several unusual features such as a two peak resistivity,
anomalous absorption feature in optical conductivity etc in the crossover
regime. We show that the proximity of the Hubbard band to the Fermi level
 is responsible for the
former while the latter happens due to optical transitions into the 
effective $f$-level. The two-band (hybridization-gap) model generally applied 
to understand the physics of these materials is shown to be inadequate in 
this regime. Qualitative agreement with pressure-dependent dc resistivity 
measurements in CeCu$_2$Si$_2$ is found while quantitative agreement with
dc transport measurements in CeRhIn$_5$, Ce$_2$Ni$_3$Si$_5$, CeFeGe$_3$
and YbIr$_2$Si$_2$ is obtained. We also obtain excellent agreement with
pressure dependent resistivity measurements in YbInAu$_2$ and the prototypical
quantum critical system CeCu$_6$. In agreement with previous observations, we find that
increasing pressure pushes Cerium materials to mixed-valence while the Yb materials cross over to the Kondo
lattice regime. Further, a remarkable agreement with 
optical conductivity experiments in YbIr$_2$Si$_2$ is obtained. We also
infer from the agreement that YbIr$_2$Si$_2$ belongs to the mixed-valent regime.
Further investigations including $d-f$ correlations in the 
periodic Anderson model are underway to understand valence transitions.

\section*{Acknowledgments}
We thank Prof.\ David E.\ Logan and Prof.\ H.\ R.\ Krishnamurthy for 
discussion and useful suggestions.
PK is grateful to CSIR, India for financial support.

\section*{Appendix}

The expression for dynamical conductivity (hypercubic lattice)
 in ~\cite{vidh05} is
\beq
\hspace{-2cm}
\sigma(\om;T)=\frac{\sigma_0}{\om} \int^\infty_{-\infty} d\om_1\,
\left[n_F(\om_1) - n_F(\om_1+\om)\right] \int^\infty_{-\infty}
d\ep\,\rho_0(\ep) D^c(\ep;\om_1) D^c(\ep;\om+\om_1)
\label{eq:a1}
\eeq
where $D^c(\ep;\om)=-\Im G^c(\ep;\om)/\pi; G^c(\ep;\om) =
\left(\gamma(\om) - \ep\right)^{-1}$ and $\gamma(\om)$ is defined below
equation~\ref{eq:hilb}.
The identity 
$
\Im z_1\Im z_2 = \Re(z_1^* z_2 - z_1z_2)/2
$ for $z_1,z_2\in \field{C}$ 
may be used to get 
\begin{displaymath}
\hspace{-2.5cm}
D^c(\ep;\om_1)D^c(\ep;\om+\om_1) =
\frac{1}{2\pi^2} \Re \left(\frac{G^{c*}(\ep;\om+\om_1) - G^c(\ep;\om_1)}
{\gamma(\om_1) - \gamma^*(\om+\om_1)} - 
\frac{G^{c}(\ep;\om+\om_1) - G^c(\ep;\om_1)}{\gamma(\om_1) - \gamma(\om+\om_1)}
\right)
\end{displaymath}
The Hilbert transform integral over $\ep$ in equation~\ref{eq:a1} may be carried
out to give the expression~ \ref{eq4}.
The $\om\rightarrow 0 $ limit of the equation~\ref{eq4} yields the dc 
conductivity. The second term involves using the L'Hopital's rule which
in turn requires the knowledge of $d G^c(\om)/d\gamma$, which we derive below. 
The c-Green's function is given by a Hilbert transform over the Gaussian
dos. The result is expressed in terms of the complementary error function
as~\cite{geor}
\beq
G^c(\om)=H[\gamma]=-is\sqrt{\pi}\exp(-\gamma^2) {\rm erfc}(-is\gamma)
\eeq
where $s=sgn(\Im\gamma)=+1$ for the retarded functions considered here.
The derivative of this with respect to $\gamma$ is straightforward and is given
by
\beq
\hspace{-2cm}
\frac{dG^c(\om)}{d\gamma} = -2\gamma G^c(\om) -is\sqrt{\pi}\exp(-\gamma^2)
(-is)\frac{d{\rm erfc}(x)}{dx}|_{x=-is\gamma} 
= 2\left(1-\gamma G^c(\om)\right)
\eeq
Using this yields equation~\ref{eq5}.

\end{document}